\documentclass{kapproc} 

\RequirePackage{graphicx}%
\RequirePackage{epsf}%
\input{psfig.sty}

\upperandlowercase
\let\footnote\savefootnote
\let\footnotetext\savefootnotetext 
\let\footnoterule\savefootnoterule 
\kluwerbib 

\def\Msolar{{M${_\odot}$\,}}

\begin{document}


\articletitle{The stellar IMF of Galactic \\clusters and its evolution}

\chaptitlerunninghead{The mass function of Galactic clusters}



\author{Guido De Marchi\altaffilmark{1}, Francesco
Paresce\altaffilmark{2} and Simon Portegies Zwart\altaffilmark{3}}

\affil{\altaffilmark{1}European Space Agency, Keplerlaan 1, 2200 AG 
Noordwijk, The Netherlands \\ 
\altaffilmark{2}European Southern Observatory, Karl-Schwarzschild-Str 
2, 85748 Garching, Germany \\
\altaffilmark{3}Instituut Anton Pannekoek, Kruislaan 403, 1098 SJ
Amsterdam, The Netherlands}

\email{gdemarchi@rssd.esa.int, fparesce@eso.org, spz@science.uva.nl}


\begin{abstract}  We show that one can obtain a good fit to the
measured main sequence mass function (MF) of a large sample of Galactic
clusters (young and old) with a tapered Salpeter power law distribution
function with an exponential truncation. The average value of the power
law index is very close to Salpeter ($\sim 2.3$), whereas the
characteristic mass is in the range $0.1-0.5$\,\Msolar and does not
seem to vary in a systematic way with the present cluster parameters
such as metal abundance and central concentration. However, a
remarkable correlation with age is seen, in that the peak mass of young
clusters increases with it. This  trend does not extend to globular
clusters, whose peak mass is firmly at $\sim 0.35$\,\Msolar. This
correlation is due to the onset of mass segregation following early
dynamical interactions in the loose cluster cores. Differences between
globular and younger clusters may depend on the initial environment of
star formation, which in turn affects their total mass.  
\end{abstract}

\section{Introduction}
 
Conflicting claims exist as to the universality of the IMF (see e.g.
Gilmore 2002) or lack thereof (e.g. Eisenhauer 2002). This
unsatisfactory state of affairs has its most likely origin in the lack
of uniformity of the experimental data used to infer the stellar IMF.
The comparison of different data-sets, obtained by different authors in
different environments (see e.g. the reviews of Scalo 1998, Kroupa 2001
and Chabrier 2003) is unfortunately hampered by systematic
uncertainties. Therefore, our only hope to assess observationally
whether the star formation process and its end result, namely the IMF,
are the same everywhere rests on our ability to secure a statistically
complete and physically homogeneous sample of stars. This is presently
possible for Galactic clusters thanks to the recent advancements in the
instrumentation (HST, VLT, etc.) and in our understanding of the
dynamical evolution of stellar systems (Meylan \& Heggie 1997). 

In Paresce \& De Marchi (2000, hereafter PDM00) we studied the
luminosity function (LFs) of a homogeneous sample of globular clusters
(GCs) and showed that, within the present uncertainties, they can all
be traced back to the same global MF and, most likely, the same IMF.
That work suggests that the latter has a log--normal form below
1\,\Msolar, with a characteristic mass $m_c=0.33 \pm 0.03$\,\Msolar and
width $\sigma = 0.34 \pm 0.04$, independently of the cluster physical
parameters or dynamical history. In a subsequent paper (De Marchi et
al. 2004) this  analysis has been extended to a homogeneous sample of
young Galactic clusters (YCs), with ages ranging from a few Myr to a
Gyr, by comparing their MF to one another and to that of GCs. Here
follows a summary of the main results.

\section{The sample}

\begin{figure}[t]
\vspace*{-0.5cm}
\hspace*{-2cm}
\centerline{\psfig{file=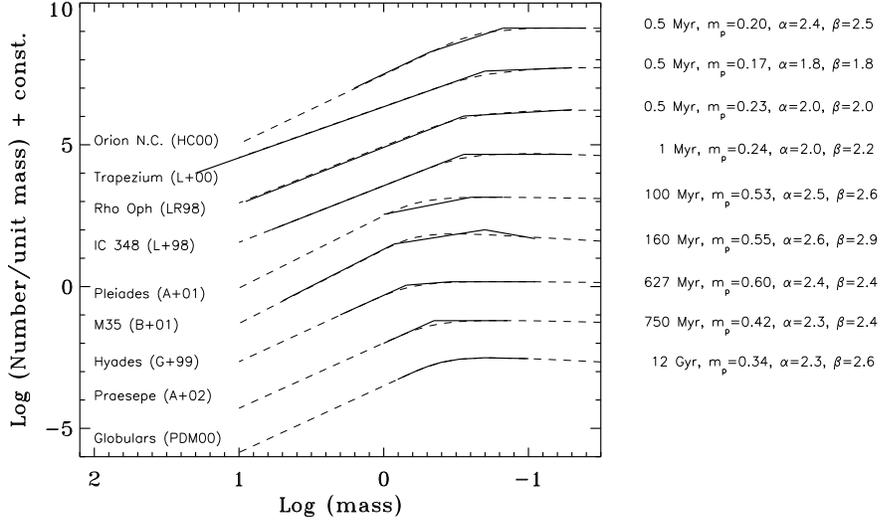,height=3in}}
\caption{Observed MFs (solid lines) are fitted with a TPL (dashed
lines). Cluster age increases from top to bottom.} 
\label{fig1} \end{figure}

While the GCs in the sample have all been observed with the same
instrument and band, and the data reduced with the same reproducible
processing (see PDM00 for details), the YCs data come from several
different sources. To enforce the highest degree of uniformity, we have
searched the literature on YC LFs with specific guidelines, namely: the
availability of recent, high quality photometry to supplement Schmidt
plate material; a clear indication of which portion of the cluster has
been studied; a solid membership selection; a reliable conversion from
magnitude to mass; a detailed explanation of any correction to the MF
to account for stellar multiplicity. The list of YCs selected in this
way and the respective references are given in Figure\,\ref{fig1}.

Since the YC data span a wide wavelength range, it is not possible to
directly compare to one another their LFs. Instead, we have
concentrated on their MFs, which most authors approximate with a
segmented power law, as done for instance by Kroupa (2001). The MFs are
shown in Figure\,\ref{fig1} (thick solid lines). Since most authors
converted magnitudes to masses using the relationships of D'Antona \&
Mazzitelli (1994), the differences in the MFs should reflect those in
the LFs.

In order to compare the MFs of the YCs in Figure\,\ref{fig1} to one
another and to that of GCs, it is useful to define some parameters that
describe the MF shape. A log--normal distribution offers a suitable
parametric description of the MF of GCs (PDM00). However, when extended
to masses above those currently observable in GCs, a log--normal MF
would fall off far more rapidly than the MF of YCs. In fact, the latter
is in most cases very close to a Salpeter power law above 1\,\Msolar.
For this reason, we have looked for a different functional form which
would accurately reproduce the observed MF of GCs and which, once
extended above 1\,\Msolar, would still be compatible with the MF of
YCs. Following the notation of Elmegreen (1999), one can write the MF
as:

\begin{equation}
f \left(m \right)=\frac{d\,N}{d\,m}\propto m^{-\alpha} \left[ 1- 
e^{(-m/m_p)^\beta} \right]
\label{eq1}
\end{equation} 

where $m_p$ is the peak mass, $\alpha$ the index of the power law
portion for high masses and $\beta$ the tapering exponent which causes
the MF to flatten at low masses. The values of the tapered power law
(TPL) parameters providing the best fit (dashed lines) to the
observations are shown on the right-hand side of Figure\,\ref{fig1},
together with the cluster age. The typical uncertainty on $m_p$ is $<
0.1$\,\Msolar.

\begin{figure}[t]
\vspace*{-1cm}
\centering
\resizebox{8cm}{!}{\includegraphics[angle=-90]{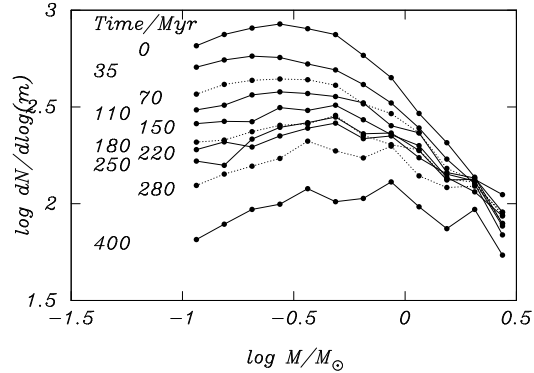}}
\caption{Temporal evolution of the MF inside the half-mass radius.}
\label{fig2}
\end{figure}

Since the index $\alpha$ has an almost negligible effect on the shape
of the MF around $m_p$, its value cannot be constrained for GCs. We
have simply assumed in this case $\alpha=2.3$ (the Salpeter value).
Space limitations do not allow us to show here the TPL fit to the MF
of each individual GC, so in Figure\,\ref{fig1} we show the average MF
(for more details, see De Marchi et al. 2004). Nevertheless, both $m_p$
and $\beta$ span a very narrow range around their average values, with
$m_p = 0.35 \pm 0.04$\,\Msolar and $\beta=2.6 \pm 0.3$ for the whole
GC sample.

\section{The evolution of the mass function} 

Figure\,\ref{fig1} reveals that the shape of the MF can change
considerably from cluster to cluster, with the peak varying widely in
mass (although $\alpha$ and $\beta$ span a range of values fully
consistent with that of GCs). The cluster MFs in Figure\,\ref{fig1} are
arranged with age increasing from top to bottom and even a casual
inspection reveals immediately a strong trend, with the MF peak
shifting to higher masses.

The most likely origin of this trend is the combined effect of mass
segregation and the limited cluster area covered by the observations.
In the absence of tidal interactions with the Galaxy, one expects the
global MF of a cluster to vary slowly with time due to evaporation. For
massive GCs this process can take several tens or hundreds of Gyr
(Gnedin \& Ostriker 1997) but, in YCs, mass segregation and the
resulting evaporation proceed more rapidly (Raboud \& Mermilliod 1998).
Portegies Zwart et al. (2001) have shown that the global MF of a
600\,Myr old cluster with a mass of 1600\,\Msolar differs only
marginally from its IMF, even when the enhanced erosion induced by the
Galactic potential is included in the calculations. However, the same
simulations show that the local MF changes dramatically in the inner
cluster regions, inside the half-mass radius. This is perfectly in line
with the YC data of Figure\,\ref{fig1}, since all the MFs shown there
were obtained in the inner cluster regions.

Without addressing here the details of a complete quantitative analysis
(see Portegies Zwart et al. 2001; De Marchi et al. 2004),
Figure\,\ref{fig2} shows the temporal evolution of the stellar MF
inside the half-mass radius of a 1600\,\Msolar model cluster. The IMF
is assumed to be that of Scalo (1986) with an initial peak at $\sim
0.4$\,\Msolar. The peak mass clearly grows with time, much in the same
way as we observe in Figure\,\ref{fig1}. Since the average stellar mass
increases towards the cluster centre, due to mass segregation, the
location of the MF peak depends steeply on the fraction of cluster area
sampled by the data: the wider the latter, the lower the peak mass.
Thus, although not specific to any one of the clusters in our sample,
the simulation shown in Figure\,\ref{fig2} proves rather convincingly
that mass segregation, combined with limited sampling of the cluster
population, can explain much of the variation noticed in
Figure\,\ref{fig1}.

\section{Conclusions}

It is, thus, likely that the YCs in Figure\,\ref{fig1} share a rather
similar IMF, since the progressive difference among their shapes grows
with age in the same sense expected from dynamical evolution. Such an
IMF would have to be very similar to the MF of the youngest clusters
($\sim 1$\,Myr old) with a peak mass $m_p \simeq 0.2$\,\Msolar or, more
likely, $m_p\simeq 0.15$\,\Msolar when account is taken of binaries
(see Kroupa 2001). The latter is very similar to the IMF of the disc
(Chabrier 2003). As discussed by PDM00, the similarity among the MF
of GCs suggests that they as well could all share the same IMF. 

An obvious question is whether the IMF is the same for both GCs and
YCs. More precisely, one could ask whether dynamical evolution in GCs
might have proceeded in such a way that the peak of their MF has moved
from an initial value of $m_p\simeq 0.15$\,\Msolar to the presently
observed $\sim 0.35$\,\Msolar. The lack of correlation between the past
dynamical history of GCs and their current global MF argues against
this hypothesis (PDM00). If, however, GCs are indeed the naked cores of
disrupted dwarf galaxies, as suggested e.g. by Martini \& Ho (2004),
one cannot exclude that their mass structure has been considerably
altered and the properties of their IMF homogenised by the stripping
process. Evidence of on-going GC disruption has been recently found in
NGC\,6712 (De Marchi et al. 1999) and NGC\,6218 (Pulone et al. 2004).
Since any low-mass stars lost by GCs should populate the halo, if the
IMF of GCs was originally similar to that of YCs, the halo MF should
also be peaked at $\sim 0.15$\,\Msolar. If, however, the MF of the halo
turns out to be similar to that currently observed in GCs, it will
indicate that their present day MF does not substantially differ from
the IMF. Unfortunately, the current uncertainties on the actual
properties of the halo MF (Graff \& Freese 1996; Gould et al. 1998) do
not presently allow us to test this hypothesis.

Regardless as to whether the IMF has a peak at $\sim 0.15$\,\Msolar or
$\sim 0.35$\,\Msolar, it appears that its functional form is well
matched by a TPL, at least for a large sample of clusters with widely
different properties. This lends support to the theoretical predictions
of Adams \& Fatuzzo (1996), Larson (1998), Elmegreen (1999; 2004) and
Bonnell et al. (2001) who suggest high- and low-mass stars form through
different processes and/or in different environments. Thus, it is
probably not premature to suggest that the difference between the peak
mass of globular and younger clusters also results from their initial
star formation environment, which in turn affects the total mass of
these systems. In spite of the many uncertainties still affecting this
investigation, the very fact that the IMF seems to have a
characteristic scale mass will hopefully soon allow us to characterise
the star formation process from the properties of the IMF itself.


\end{document}